# Suspension of the fiber mode-cleaner launcher and measurement of the high extinction-ratio ($10^{-9}$) ellipsometer for the Q & A experiment

**Hsien-Hao Mei\*, Sheng-Jui Chen, and Wei-Tou Ni**

Center for Gravitation and Cosmology, Department of Physics,
National Tsing Hua University, Hsinchu, Taiwan 30013, Republic of China

\*E-mail: d887309@phys.nthu.edu.tw

**Abstract.** The Q & A experiment, first proposed and started in 1994, provides a feasible way of exploring the quantum vacuum through the detection of vacuum birefringence effect generated by QED loop diagram and the detection of the polarization rotation effect generated by photon-interacting (pseudo-)scalar particles. Three main parts of the experiment are: (1) Optics System (including associated Electronic System) based on a suspended 3.5-m high finesse Fabry-Perot cavity, (2) Ellipsometer using ultra-high extinction-ratio polarizer and analyzer, and (3) Magnetic Field Modulation System for generating the birefringence and the polarization rotation effect. In 2002, the Q & A experiment achieved the Phase I sensitivity goal. During Phase II, we set (i) to improve the control system of the cavity mirrors for suppressing the relative motion noise, (ii) to enhance the birefringence signal by setting-up a 60-cm long 2.3 T transverse permanent magnet rotatable to 10 rev/s, (iii) to reduce geometrical noise by inserting a suspended polarization-maintaining optical fiber (PM fiber) as a mode cleaner, and (iv) to use ultra-high extinction-ratio ($10^{-9}$) polarizer and analyzer for ellipsometry. Here we report on (iii) & (iv); specifically, we present the properties of the PM-fiber mode-cleaner, the transfer function of its suspension system, and the result of our measurement of high extinction-ratio polarizer and analyzer.

PACS numbers: 04.80.-y, 12.20.-m, 14.80.Mz, 07.60.Ly, 07.60.Fs, 33.55.Ad

## 1. Introduction

Birefringence of vacuum is predicted by QED (Quantum Electrodynamics) in external transverse electrical or magnetic field (**E** or **B**) [1-6]. For gaseous matter, the birefringence phenomenon is known as Cotton-Mouton effect. The velocity of light ($\propto 1/n$) propagating through a transverse **B** field depends on **B**$^2$ and the orientation of polarization of light relative to **B**. For $B = |\mathbf{B}| = 2.5$ T, the corresponding vacuum birefringence would be $\Delta n = n_\parallel - n_\perp = 2.5 \times 10^{-23}$ according to QED. Experimental schemes for detecting vacuum birefringence were proposed by Iacopini and Zavattini [5], and Ni, Tsubono, Mio, Narihara, Chen, King and Pan [6]. First experiment was done in 1993 by Cameron *et al.* [7] with good upper bounds on vacuum birefringence and polarization rotation. A pseudo-scalar interaction with electromagnetism ($L_1 \sim \phi F_{ij} F_{kl} e^{ijkl}$) was proposed which was empirically allowed in the study of equivalence principles [8-10]; in terms of Feynman diagram, this interaction gives a 2-photon-pseudo-scalar vertex. With this interaction, vacuum becomes birefringent and dichroic [11-13]. In 1994, 3experiments were put forward and started for measuring the vacuum birefringence: the PVLAS experiment [14], the Fermilab P-877 experiment [15], and the Q & A (QED & Axion) experiment [16]; these experiment were reported in the "Frontier Test of QED and Physics of Vacuum" Meeting in 1998. Fermilab P-877 experiment was terminated in 2000. This year in the QED-2005 (QED, Quantum Vacuum and the Search for New Forces, Les Houches, France, June, 2005) conference, again, 3 experiments were reported: the PVLAS experiment [17], the BMV experiment [18], and the Q & A experiment [19]. A compilation of basic characteristics of these 3 experiments were given in [20]. All 3 experiments use high-finesse Fabry-Perot Interferometer (FPI) cavity to enhance the effect to be measured. The PVLAS experiment reported a positive measurement of polarization rotation and suggested a possible interpretation of this result to the existence of a pseudoscalar particle coupled to photons [17, 21].

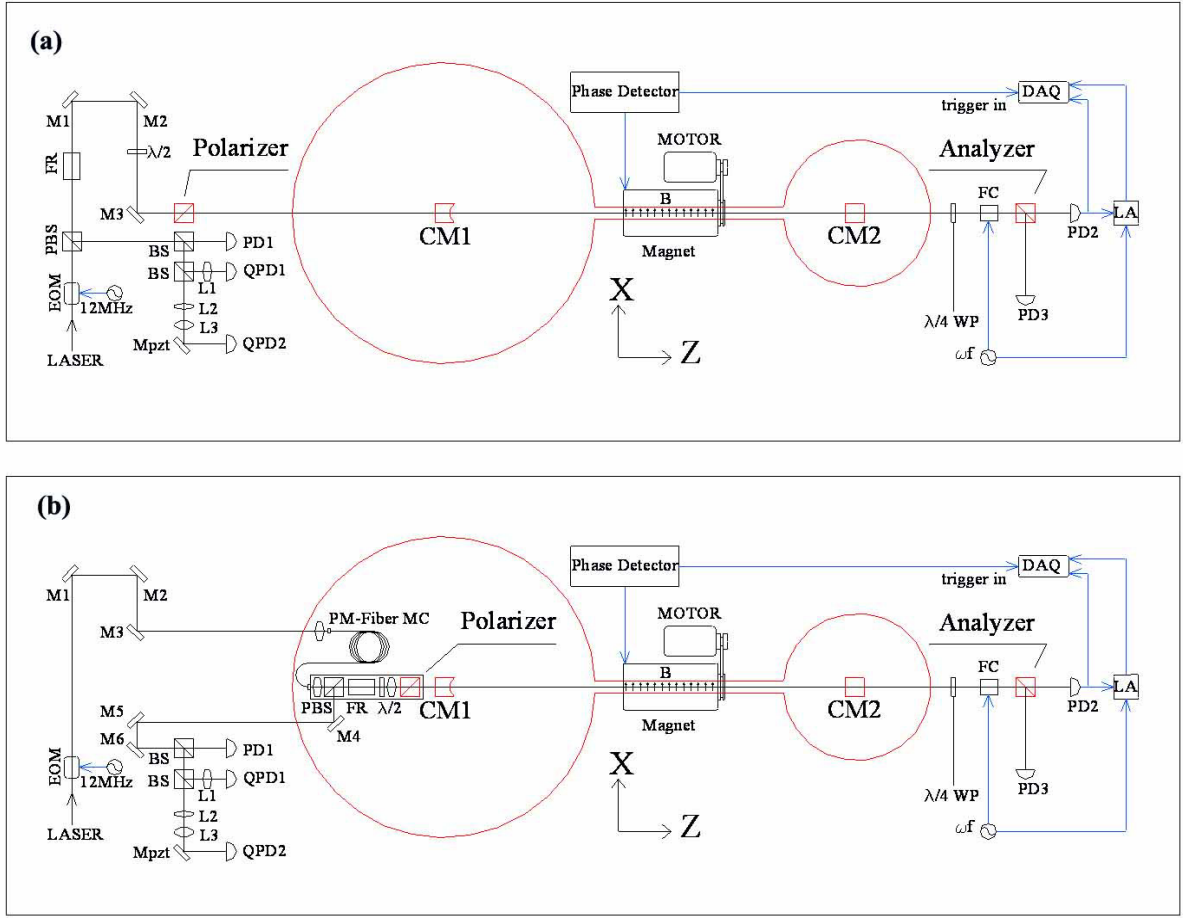

**Figure 1.** The experimental set-up for the Phase II design of the Q & A experiment. **(a)** Current status. EOM: electo-optical modulator. M1 ~ M6: mirrors. L1 ~ L3: lens. (P)BS: (polarizing) beam splitter. (Q)PDs: (quadrant) photo-detectors. MC: mode cleaner. FR: Faraday rotator. λ/2: half-wave plate. λ/4: quarter-wave plate. CM1, 2: Fabry-Perot Interferometer (FPI) cavity mirrors. FC: Faraday cell. LA: lock-in amplifier. DAQ: data acquisition. The birefringence signal was modulated by the rotating permanent magnet (up to 10 rev/s) and enhanced by the suspended Fabry-Perot cavity. With a quarter-wave plate and additional modulation by the Faraday cell, the birefringence signal can be extracted as a polarization rotation signal through a lock-in amplifier with precisely detected signal phase for long-term integration when automatic alignment control loop is closed. **(b)** Design of the Q & A experiment with fiber-mode cleaner inserted.

Our experimental scheme is shown in figure 1. We had built a 3.5 m prototype X-pendulum-double-pendulum suspended FPI with Finesse $F = 11,620$ and a slow amplitude modulatable (1 Tesla at 50 mHz) bi-polar electrical magnet of length $L_B = 23$ cm, and achieved the Phase I ellipticity detection sensitivity goal [22] in 2002. After 2002, we are working on improving this prototype by implementing an automatic alignment system for suspension control, by using a 60 cm 2.3 T rotatable (DC-10 rev/s) permanent magnet to enhance signal, and by employing a pair of high extinction-ratio polarizer and analyzer for ellipticity/polarization rotation detection (figure 1(a)). We plan to insert a PM fiber (polarization-maintaining fiber) as a mode cleaner with its output port suspended by a double pendulum before the laser light entering the FPI and integrate the feed-back control mechanism into the current automatic alignment system (figure 1(b)).

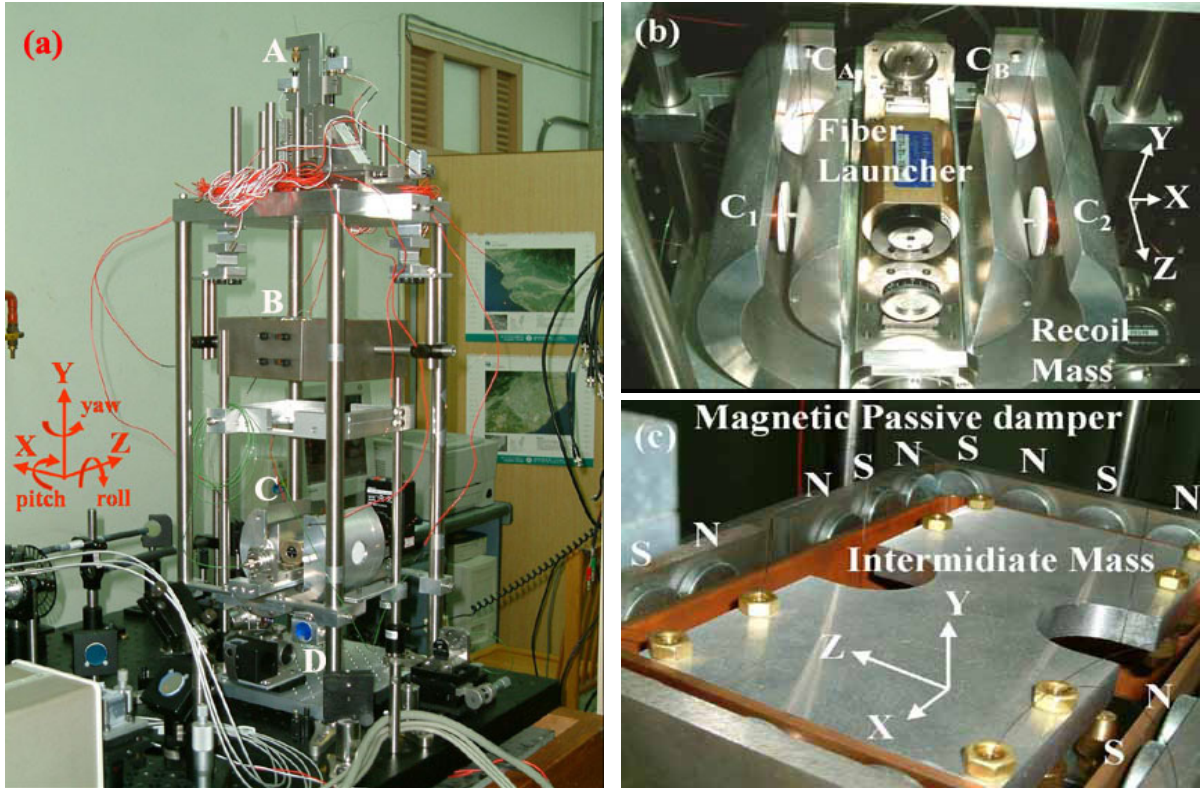

**Figure 2. (a)** Double-pendulum suspension for the fiber launcher holding the fiber output port and related optics. The coordinate system is identical to that in figure 1. **A**: suspension point. **B**: intermediate mass (first stage) with passive damper. **C**: fiber launcher (final stage) with recoil mass. **D**: HP 3-axis heterodyne interferometer and end mirror attached on a very light mass block connected rigidly to the fiber launcher through a hole in the bottom of the recoil mass. **(b)** A closer look of the final stage, recoil mass and the coil-magnet actuators. The coils ($C_1$, $C_2$, $C_A$, $C_B$) are mounted on the recoil mass. **(c)** Intermediate mass and chess-board organized magnetic damping.

## 2. Fiber mode cleaner launcher and its suspension system

Optical fiber as mode cleaner has been adopted in the experiment as mentioned in [23-24]. The fundamental mode emitted from a single mode PM fiber is regarded as a good approximation of the FPI's fundamental $TEM_{00}$ mode. For further reduction of seismic noise in our detection band (10-20 Hz), it is obvious that we need a suspension solution for the fiber output coupler. In figure 2(a) and 2(b), we made a suspension system for this output port of the PM fiber, together with a fiber table, a PBS for retrieving the Pound-Drever-Hall locking signal, a Faraday rotator, a half-wave-plate, and the polarizer as shown in figure 1(b). The whole assembly is named "the fiber launcher" and its position monitored by an HP 3 axis heterodyne interferometer (see figure 2(a), part **D**). A recoil mass with weight equal to the fiber launcher was chosen to keep the center-of-mass stable and prevent the re-injection of excess noise from the control loop to the common upper stage (figure 2(c)). Here we show the measurements of the translational and rotational noise spectrum densities of the stage (figure 3(a), 3(b)). The main resonances were attenuated by 10-20 dB.

Some open loop transfer functions (TFs) of the voltage-controlled current signal in the coil-magnet actuators ($C_1$, $C_2$, $C_A$, $C_B$) to the displacement of the final stage monitored by the HP heterodyne interferometer were also measured and shown in figure 4. The TFs for each individual coil were measured with a periodic linear frequency-sweep (0.1-10 Hz) input current signal. The input signal and the output displacements of the stage were simultaneously recorded by an industrial PC at a sampling rate of 85 Hz. The data was then off-line analyzed. The $X/V_{in}$ and $Z/V_{in}$ TFs could be closely simplified to simple pendulum formula due to the quiet center-of-mass motion after strong passive damping. However, weight matching does not imply or guarantee momentum-of-inertia matching; the TFs for yaw/$V_{in}$ still looks like a double pendulum. Further tests will be proceeded after completing the control servo design from the measured open loop TFs of this suspension system.

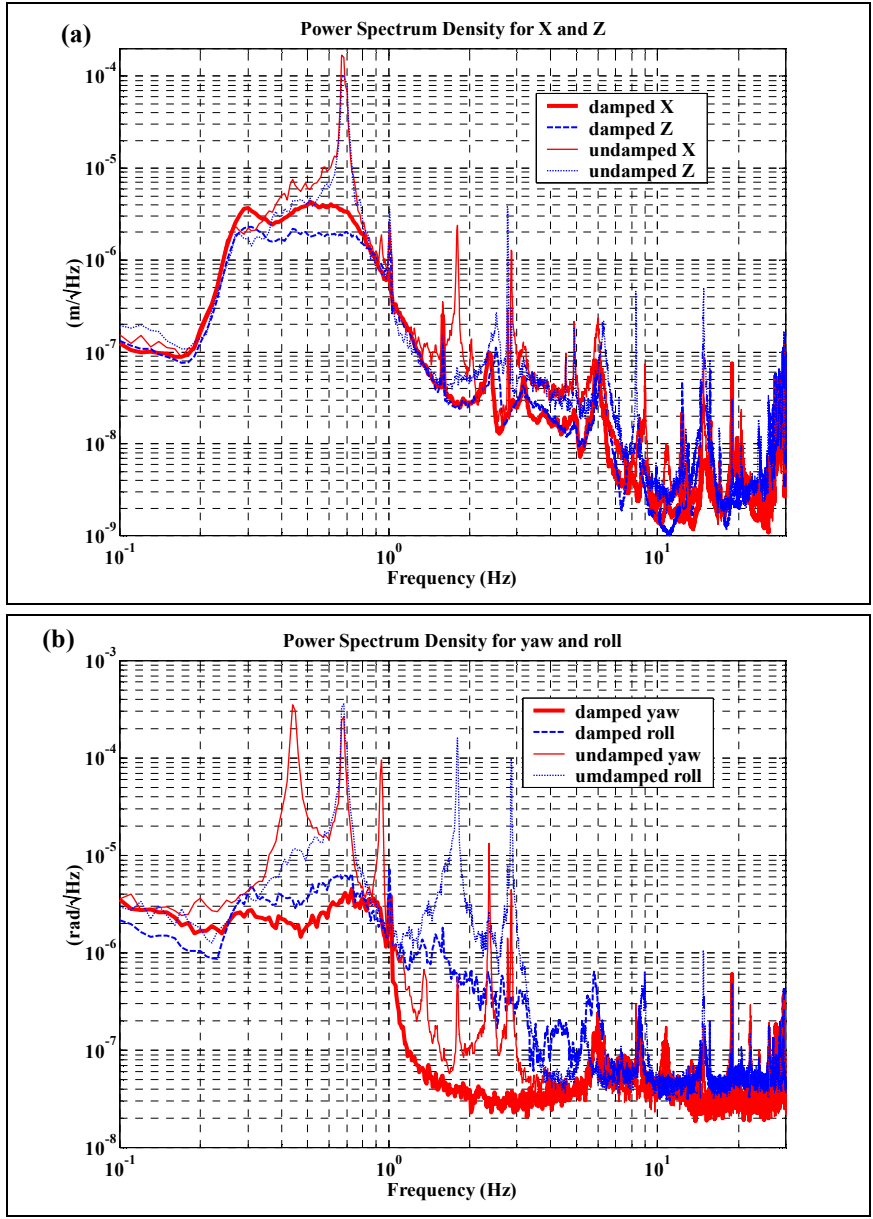

**Figure 3. (a)** Power spectrum of displacements of the final stage in the X and Z directions with/without the passive damper. **(b)** Power spectrum of the rotations of the final stage in the yaw and roll directions defined in figure 2.

### 3. High extinction ratio polarizer for the ellipsometer

For precise determination of the birefringence (the ellipticity) signal, it is necessary to understand the detection limit of the ellipsometer. The ellipsometer can be simply characterized by the extinction ratio ($\sigma^2$) of the analyzer from the Malus' Law [25]:

$$I = I_0 \, (\sigma^2 + \sin^2 \alpha) \approx I_0 \, (\sigma^2 + \alpha^2)$$

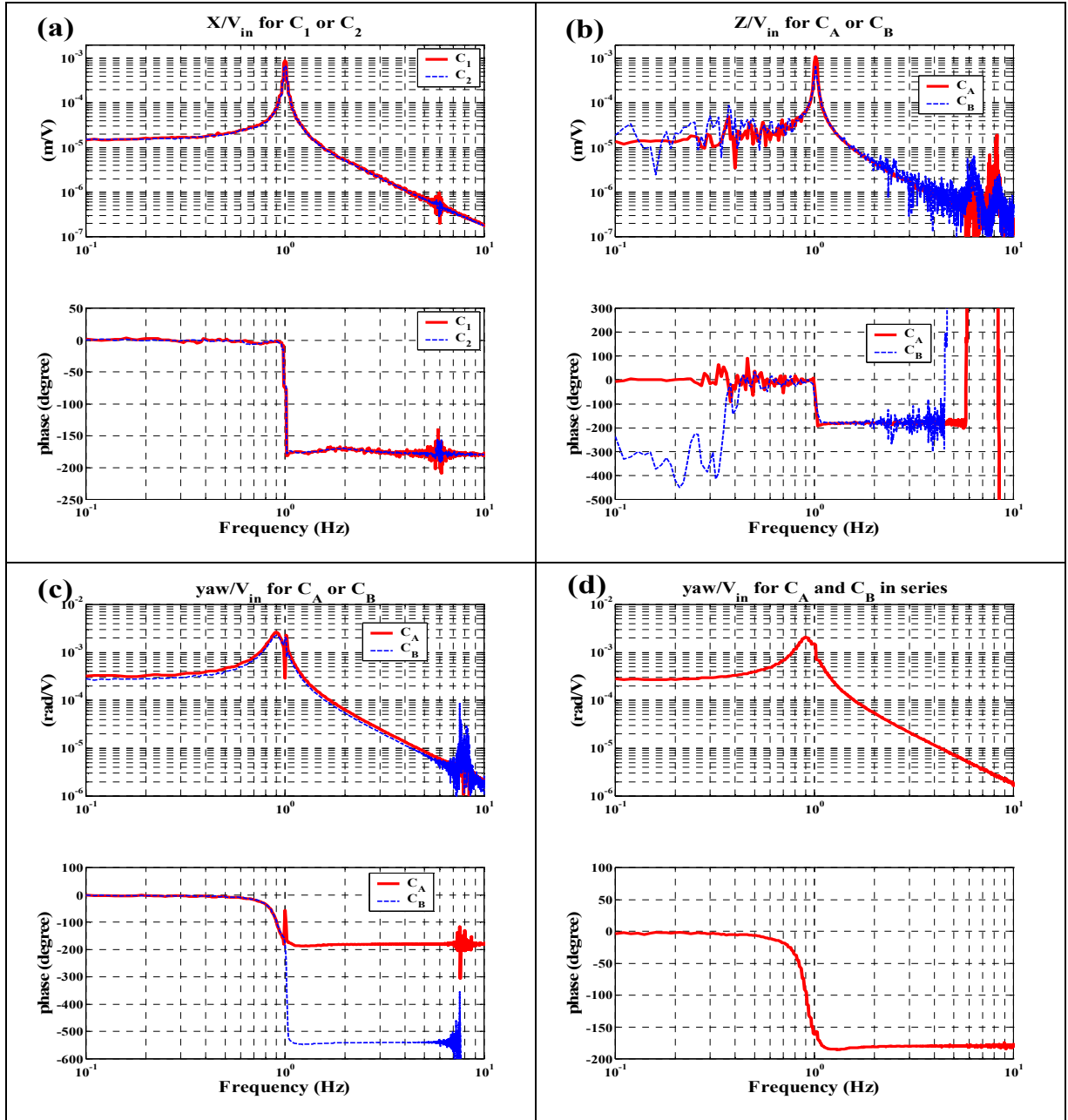

**Figure 4(a)-(d).** Transfer functions of the voltage control signal of the coils ($C_1$, $C_2$, $C_A$, $C_B$, indicated in figure 2(b)) to the displacement of the suspended final stage with respect to the HP interferometer. The main transfer functions in directions of X, Z, and yaw are shown. The transfer functions in translational degrees of freedom (X, Z) are of the form of simple pendulums due to weight-matching, while in rotational degree of freedom (yaw) it is of the form of a double pendulum due to non-matching of momentum-of-inertia.

where $I_0$ and $I$ are the measured intensity of light before and after it passes through a pair of crossed polarizer and analyzer, and $\alpha$ is the misaligned angle between the crossed polarizer and analyzer. To experimentally determine $\sigma^2$, we focus in the region where $\alpha$ is most sensitive to $\sigma^2$ to obtain a good fitting. This demands $\alpha^2$ to be comparable to $\sigma^2$ and the measurement range of $\alpha$ be larger than $2\sigma$. For $\sigma^2 \sim 10^{-9}$, we use a measurement range larger than $10^{-4}$ rad and angular resolution less than $10^{-6}$ rad. We use an HP 3-axis heterodyne interferometer to measure the distance and angle. For rotating the analyzer, we employ a combination of piezo transducer (PZT) and flexure structure as in [26]. The resolution of heterodyne interferometry for length metrology [27] can be readily converted into the resolution of angle measurement. The angular resolution is $3.5 \times 10^{-7}$ rad.

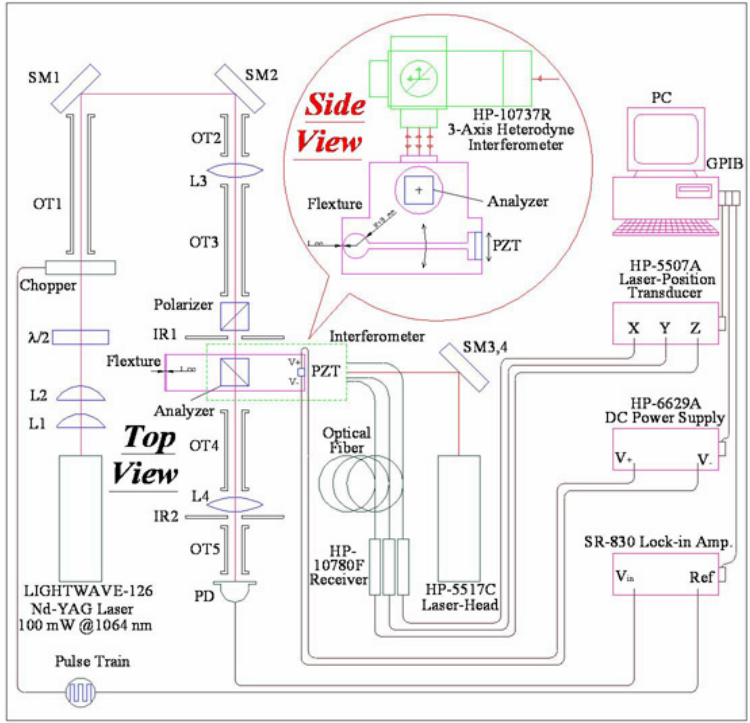
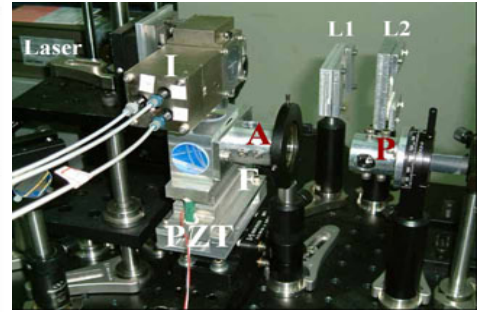

**Figure 5. Left**: The scheme for measuring extinction ratio of 3 different analyzers. L1 ~ L4: lens. λ/2: half wave plate. SM1 ~ SM4: steering mirrors. OT1 ~ OT5: opaque tubes. IR1, 2: iris. PD: photo detector. PZT: piezo electric transducer. The analyzer was fixed to a piezo-controlled angular rotator with flexure structure as an assembly, and an HP 3-Axis heterodyne interferometer was used to resolve the small angle change. The side view of them is shown in the circle inset. The angle was detected with a resolution of $3.5 \times 10^{-7}$ rad. **Top**: A photo shows the spatial arrangements for the polarizer **P**, the analyzer **A**, the heterodyne interferometer **I**, the piezo **PZT**. The 1 mm thick flexure **F** is beyond our sight. L1, L2 and the Laser also appear as background.

The experimental set up is shown in figure 5: A well collimated Nd-YAG laser beam is intensity-modulated by a chopper before passing through a pair of crossed polarizer and analyzer. A photo-detector lock-in detects the transmitted laser power with the chopper signal while the heterodyne interferometer detects the angular position change induced by the voltage-driven PZT. The result of a typical measurement is shown in figure 6(a). Each data point in the figure is the average of 100 times data-taking both in angle and in intensity. The standard deviation of the intensity measurement is represented by an error bar; the standard deviation of the angle measurement is not shown for clarity of drawing. We estimate the extinction ratio of the analyzer by first recording these average values and error bars of independent angular sweeps. Then for each independent sweep, we use Monte-Carlo method to generate 10,000 samples consistent with average values and error bars and least-square-fit these samples to obtain a distribution of parabolic parameters. From this distribution, the extinction ratio $\sigma^2$ and its standard deviation are determined. For 3 different analyzers with identification numbers 3, 4, and 5, the experimental extinction ratio and error bar for all independent sweeps are shown in figure 6(b). The extinction ratios were in the order of $10^{-9}$ to $10^{-10}$. The highest extinction value measured was $(5.21 \pm 0.41) \times 10^{-10}$, which we believe to be the highest value so far that has ever been reported at 1064 nm wavelength. For 633 nm wavelength, Takubo *et al*. has measured an extinction value of $(2.9 \pm 0.2) \times 10^{-10}$ in 1998 [26].

## 4. Discussion and outlook

From the Malus' equation, it is obvious that higher extinction ratio would give better resolution/ sensitivity in ellipticity/polarization rotation detection, which related to the birefringence signal $\Delta n$ by

$$\Psi = 2FL_B\Delta n / \lambda$$

where $\Psi$ is the ellipticity, $F$ is the finesse of the FPI, and $\lambda$ is the wave-length of laser. Hence the corresponding ellipticity signal for vacuum birefringence under the Q & A phase III condition that magnetic field $B = 2.5$ Tesla, $F = 100{,}000$ and $L_B = 5$ m is $2.5 \times 10^{-11}$ rad. For the current Q & A experiment, the sensitivity of ellipticity detection at the frequency around 20 Hz is $2 \times 10^{-6}$ rad·Hz$^{-1/2}$.

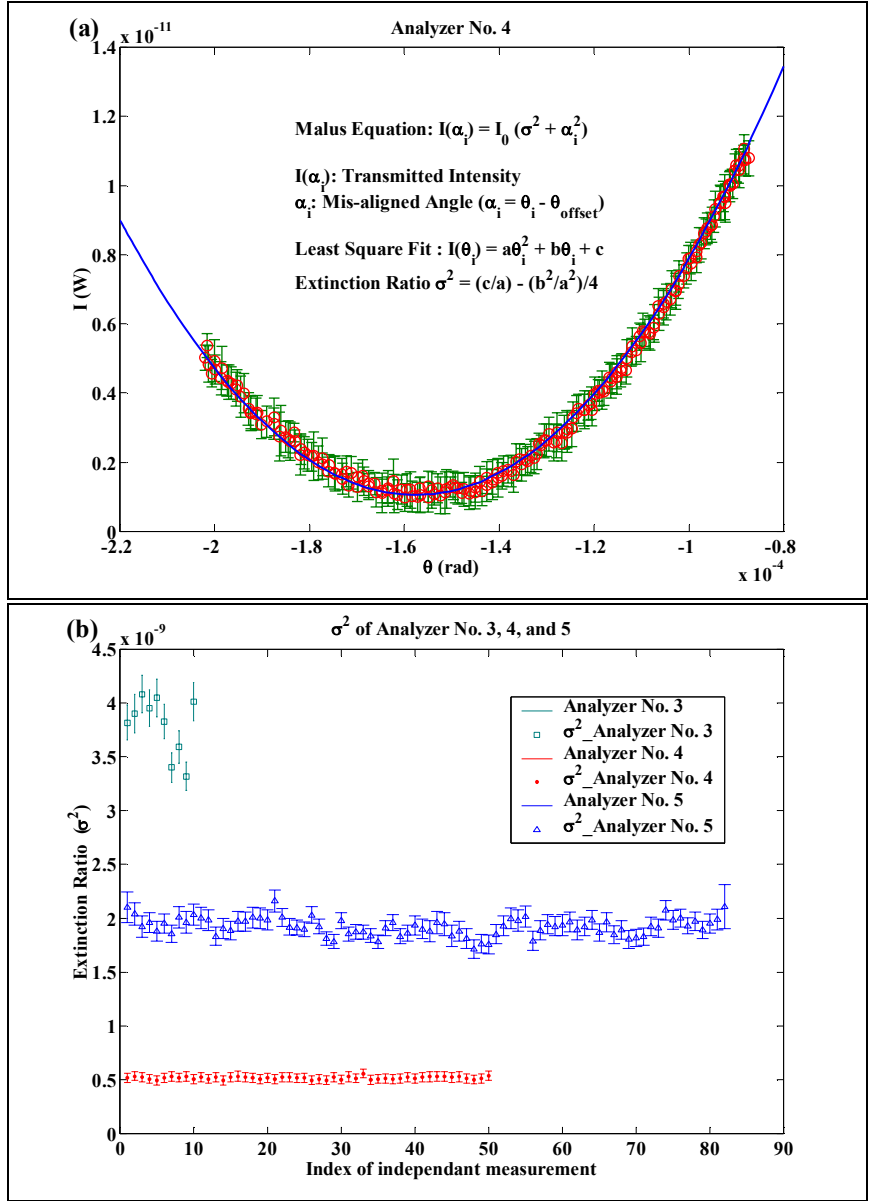

**Figure 6.** Experimental measurement data and the result for the extinction ratio of analyzers. **(a)** The measured transmitted power v.s. misalignment angle obeys the Malus equation. The error bar for angle is not shown here for the clarity of drawing. **(b)** The measured results and error bars of the extinction ratio of analyzers. The highest extinction ratio measured is $(5.21 \pm 0.41) \times 10^{-10}$.

The sensitivity goal is $5 \times 10^{-8}$ rad·Hz$^{-1/2}$ at 10-20 Hz frequency band (for Phase II). In the near future we can perform a high-resolution determination of gaseous birefringent effect known as Cotton-Mouton effect for various kinds of gases such as $N_2$, $O_2$, $H_2$, $CO_2$ or He at 1064 nm wavelength under different pressure which have not been measured or reported before [28-29]. In our next step, we plan to stabilize the CM1 and CM2 of the FPI (see figure 1) in the X, Z, and yaw (figure 1, 2) directions through a feedback control system. The system mainly consists of HP heterodyne interferometers and a VME-bus based real-time control system. We believe such stabilization of the CM1 and CM2 can reduce the rms displacement noise and alleviate the heavy load in the existing control loops. It also helps to reduce the roll motion (figure 1, 2) excited mostly by the arc-like motion of pendulum, as well as the possibility of the swing of laser beam spot on the mirror surface. These can all accumulate small ellipticity or polarization change and appears as noises in ellipticity measurement. The fast rotating magnet (up to 10 rev/s) is well shielded [30]. The measured magnetic field at a distance 1 cm away from main body along its rotating axis (i.e., the beam axis), is 1 Gauss, while the CM1 and CM2 are over 1 m away from it. The possibility of disturbing the cavity mirrors through magnetic coupling of this field to the actuators is considered to be negligible.

We are also looking forward to performing a measurement of polarization rotation to compare with the PVLAS result [17, 21] after the improvement in this paper and in the companion paper [31].

## 5. Acknowledgment
We thank the National Science Council for supporting this program (NSC 93-2112-M-007-022, NSC 94-2112-M-007-012). We also thank the referees for helpful comments on the presentation of this paper.